\documentclass[12pt]{article}
\usepackage{a4wide}
\usepackage{graphicx}
\usepackage{amsmath}
\usepackage{slashed}
\usepackage{psfrag}
\usepackage{color}

\begin{document}
{\renewcommand{\thefootnote}{\fnsymbol{footnote}}
\medskip
\begin{center}
{\LARGE Loop quantum gravity as an effective theory}\\
\vspace{1.5em}
Martin Bojowald\footnote{e-mail address: {\tt bojowald@gravity.psu.edu}}\\
\vspace{0.5em}
Institute for Gravitation and the Cosmos,\\
The Pennsylvania State
University,\\
104 Davey Lab, University Park, PA 16802, USA\\
\vspace{1.5em}
\end{center}
}

\setcounter{footnote}{0}

\newcommand{\lP}{\ell_{\rm P}}
\newcommand{\md}{{\mathrm d}}
\newcommand{\tr}{{\mathrm{tr}}}
\newcommand{\vt}{\vartheta}
\newcommand{\vp}{\varphi}

\begin{abstract}
  As a canonical and generally covariant gauge theory, loop quantum gravity
  requires special techniques to derive effective actions or equations. If the
  proper constructions are taken into account, the theory, in spite of
  considerable ambiguities at the dynamical level, allows for a meaningful
  phenomenology to be developed, by which it becomes falsifiable. The
  tradiational problems plaguing canonical quantum-gravity theories, such as
  the anomaly issue or the problem of time, can be overcome or are irrelevant
  at the effective level, resulting in consistent means of physical
  evaluations. This contribution presents aspects of canonical equations and
  related notions of (deformed) space-time structures and discusses
  implications in loop quantum gravity, such as signature change at high
  density from holonomy corrections, and falsifiability thanks to
  inverse-triad corrections.
\end{abstract}

\section{Introduction}

Loop quantum gravity \cite{Rov,ThomasRev,ALRev} is a proposal for a canonical
quantization of general relativity. By a careful use of basic variables
suitable for a quantum representation independent of auxiliary metric or
causal structures, it has shed light on several aspects of the quantum
geometry of space. The dynamics of the theory, however, remains poorly
controlled, and therefore it is not clear what structure of quantum space-time
it implies. Dynamical operators are subject to quantum ambiguities, and their
evaluation is still plagued by long-standing conceptual problems of canonical
quantum gravity, most famously the problem of time
\cite{KucharTime,Isham:Time,AndersonTime}.

That much-needed progress especially on the last-mentioned problem is lacking
is illustrated for instance by the proliferating use of ``deparameterized''
quantum theories, in which the time variable is fixed once and for all (as a
matter degree of freedom rather than a coordinate). The space-time gauge is
partially fixed, even though one set out to derive properties of quantum
space-time.  Such time-fixed systems have been seriously proposed not just as
toy models but even as complete quantum theories of gravity (see e.g.\
\cite{DeparamQG}). With these attempts, one no longer aims at gaining control
over quantum space-time and its physical effects, for which one would have to
show that quantum observables are independent of the choice of time, be it an
internal matter clock or a coordinate; no such results of independence have
been provided in deparameterized models.

With problems like these still outstanding, it remains questionable whether
the theory can be considered as fundamental. (Of course, this is not to say
that the theory could not achieve fundamental status with further, significant
progress.)  Nevertheless, the theory's results regarding the quantum geometry
of space and background independence may still be of physical interest,
provided they can manifest themselves in sufficiently characteristic ways on
the typical scales of gravitational phenomena, far removed from the
microscopic Planck scale. This is the realm of effective theory, a powerful
viewpoint in many examples (not only) of high-energy phenomena, so also in
loop quantum gravity as laid out in this contribution.

The dynamical problems of loop quantum gravity are not specific to this
particular approach but have a general origin in relativistic properties of
space-time.  Generally covariant theories, such as general relativity, have
complicated gauge structures (coordinate transformations, or hypersurface
deformations) that can best be addressed with canonical techniques. (For
canonical methods in gravity, see \cite{CUP}.) In such settings they show
their troubling face most directly, but they also sneak through other
approaches, for instance those using path integrals or spin foams where
finding the correct integration measure is a related problem.  These issues
require special care, new methods of quantum field theory and semiclassical
or effective descriptions. Here we present an overview of the following
techniques, suitable for physical evaluations of the theory: (i) A canonical
derivation of effective equations for quantum dynamics, and (ii) a discussion
of general covariance in quantum gravity and deformed space-time
structures. By bringing these parts together, an effective theory of loop
quantum gravity is obtained.

\section{Effective theories}

In quantum theory, every local classical degree of freedom, $(q,p)$ in
canonical form, amounts to infinitely many quantum degrees of freedom ---
expectation values $\langle\hat{q}\rangle$, $\langle\hat{p}\rangle$,
fluctuations (squared) $\langle\hat{q}^2\rangle-\langle\hat{q}\rangle^2$, and
so on with higher powers. An effective theory, in general terms, aims to
describe some properties of quantum theory by interactions of
finitely many local degrees of freedom.  For instance, if we keep only the
expectation values, we have the classical limit.  Expectation values together
with fluctuations, taking values near saturation of the uncertainty relation,
provide the first-order approximation in $\hbar$.  The higher the
$\hbar$-order, the more degrees of freedom must be considered (related but not
identical to higher time derivatives).  In a formal limit including the order
$\hbar^{\infty}$, we are back at quantum theory, but in perturbative
guise. Some subtle effects may be missed, for instance different inequivalent
self-adjoint extensions of Hamiltonians. But many dynamical properties
independent of subtleties can be derived conveniently using effective
theory. One may therefore hope that some of the technical and conceptual
problems of quantum gravity or quantum cosmology can be simplified as well,
and this hope is indeed borne out.

\subsection{Quantum phase space}

Like most other questions, the idea of effective theories can easily be
illustrated by the harmonic oscillator, with a quadratic Hamiltonian
$\hat{H}=\frac{1}{2m}\hat{p}^2+ \frac{1}{2}m\omega^2\hat{q}^2$ for canonical
commutation relations $[\hat{q},\hat{p}]=i\hbar$. (However,
harmonic-oscillator results should be taken with a grain of salt when it
comes to general behavior, as we will see in the present context as well.)
For an effective theory in terms of expectation values, we compute equations
of motion
\[
 \frac{{\rm d}}{{\rm d} t}\langle\hat{q}\rangle=
\frac{1}{i\hbar}\langle[\hat{q},\hat{H}]\rangle= 
\frac{1}{m}
\langle\hat{p}\rangle \quad,\quad
 \frac{{\rm d}}{{\rm d} t}\langle\hat{p}\rangle =
\frac{1}{i\hbar}\langle[\hat{p},\hat{H}]\rangle
 =-m\omega^2\langle\hat{q}\rangle
\]
which can be solved directly. Although they amount to just the classical
equations for $q$ and $p$, their solutions determine exact quantum properties.

To go beyond the classical order free of $\hbar$-dependent terms, we can use
the same type of equations of motion to derive dynamical laws for fluctuations
$(\Delta O)^2=\langle\hat{O}^2\rangle-\langle\hat{O}\rangle^2$ of $q$ and $p$
and, as it turns out to be necessary, the covariance.
$C_{qp}=\frac{1}{2}\langle\hat{q}\hat{p}+\hat{p}\hat{q}\rangle-
\langle\hat{q}\rangle\langle\hat{p}\rangle$. These variables provide all
degrees of freedom to second order, with expectation values of quadratic
functions of the basic operators. In a semiclassical state, the values are
of the order $\hbar$, as one can verify explicitly for a Gaussian.

Dynamically, we have the equations
\begin{eqnarray}
 \frac{{\rm d}}{{\rm d}t} (\Delta q)^2&=&
 \frac{\langle[\hat{q}^2,\hat{H}]\rangle}{i\hbar}-
2\langle\hat{q}\rangle\frac{{\rm d}\langle\hat{q}\rangle}{{\rm d}t}= 
\frac{2}{m}C_{qp}\label{dqq}\\
\frac{{\rm d}}{{\rm d}t} C_{qp} &=& -m\omega^2(\Delta
q)^2+\frac{1}{m}(\Delta p)^2 \label{dCqp}\\
\frac{{\rm d}}{{\rm d}t} (\Delta p)^2&=& -2m\omega^2C_{qp} \label{dpp}
\end{eqnarray}
and we should also restrict the variables for them to correspond to a true
state: they are
subject to the (generalized) uncertainty relation
\begin{equation}\label{Uncert}
 (\Delta q)^2(\Delta p)^2-C_{qp}^2\geq \frac{\hbar^2}{4}\,.
\end{equation}

\begin{figure}
  \includegraphics[height=.3\textheight]{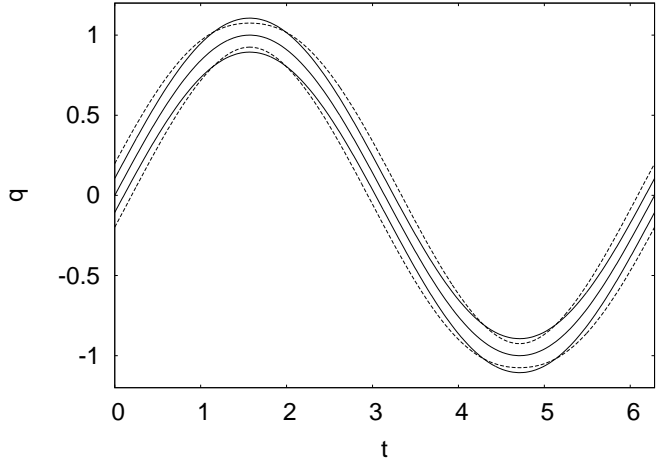}
  \caption{Dynamical coherent states of the harmonic oscillator, uncorrelated
    with constant fluctuations (solid) or correlated with oscillating
    fluctuations (dashed). All states spread around the same time-dependent
    expectation value $q(t)$ (central line) but differ in the dynamics of
    their position fluctuations (lines around the mean). \label{Squeezed}}
\end{figure}

Just like expectation-value equations, these second-order equations are linear
and can easily be solved, providing non-classical information about quantum
states. Stationary states, for instance, require $C_{qp}=0$ for the variables
to remain constant in time, together with $\Delta p=m\omega\Delta q$ from
(\ref{dCqp}). They satisfy the uncertainty relation if $\Delta q\geq
\sqrt{\hbar/2m\omega}$. The uncertainty relation is saturated for $\Delta
q=\sqrt{\hbar/2m\omega}$, in which we find the correct fluctuations of the
harmonic-oscillator ground state. More general squeezed coherent states with
$C_{qp}\not=0$, still saturating the generalized uncertainty relation, have
time-dependent fluctuations, such as a spread oscillating oscillating with
frequency $2\omega$; see Fig.~\ref{Squeezed}. One can explicitly solve
(\ref{dqq})--(\ref{dpp}) for the position fluctuation
\begin{equation}
 (\Delta q)^2(t)= \frac{C_0}{m\omega} \sin(2\omega t)+
 \frac{1}{2}(Q_0-P_0/m^2\omega^2) \cos(2\omega t)+
 \frac{1}{2}(Q_0+P_0/m^2\omega^2)
\end{equation}
with $C_0$, $Q_0$ and $P_0$ the initial values of $C_{qp}$, $(\Delta q)^2$ and
$(\Delta p)^2$, respectively. With similar solutions for $(\Delta p)^2(t)$ and
$C_{qp}(t)$, one sees that $(\Delta q)^2(\Delta p)^2-C_{qp}^2$ is constant:
the states are dynamical coherent states.

\begin{figure}
  \includegraphics[height=.43\textheight]{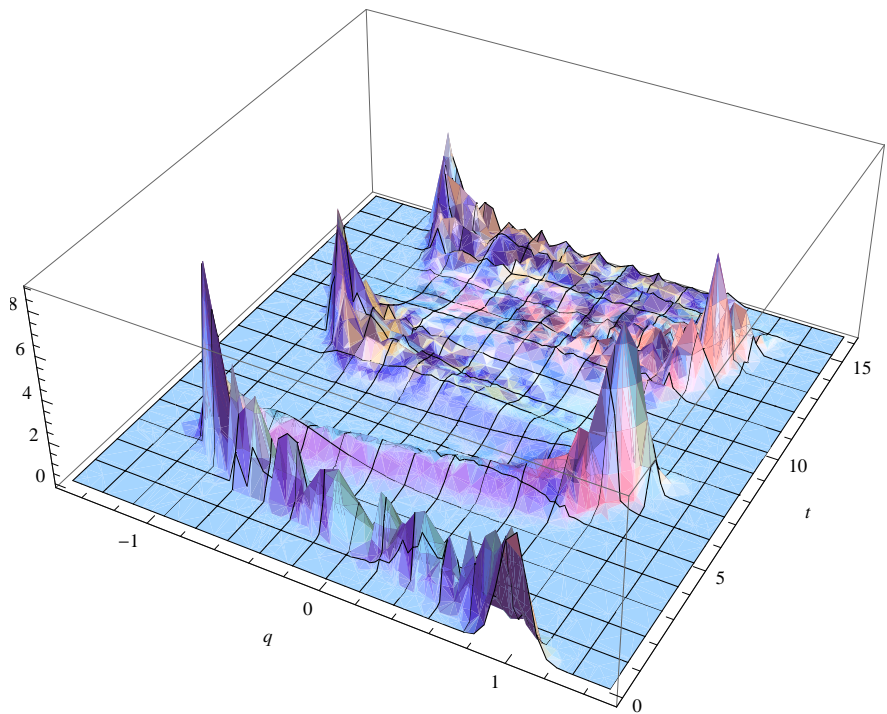}
  \caption{A wave function subject to the dynamics of a ``relativistic''
    harmonic oscillator, with quantum back-reaction. Starting as a Gaussian
    (front), the state for some time follows the classical oscillating
    trajectory but spreads out and eventually shows no clear trajectory
    \cite{EffConsComp}.\label{RelOscWave}}
\end{figure}

In this system, exact quantum properties follow from finitely many
variables. A more complicated but still tractable model is the
``relativistic'' harmonic oscillator, in which the energy
$\hat{E}^2=\frac{1}{2m}\hat{p}^2+ \frac{1}{2}m\omega^2\hat{q}^2$, compared to
the standard harmonic oscillator, enters quadratically \cite{EffConsComp}. As
Fig.~\ref{RelOscWave} shows, simple properties of an initial coherent state
become much more complicated as time goes on, and deviations from classical
trajectories occur. This form of quantum back-reaction --- the influence of
the shape of a state on the trajectory --- happens generically in quantum
systems, while the strict decoupling of quantum variables and expectation
values is special to the harmonic oscillator and a few other systems.

For controlled deviations from harmonicity, we look at an anharmonic
oscillator with non-quadratic Hamiltonian
\[
 \hat{H}=\frac{1}{2m}\hat{p}^2+V(\hat{q})= \frac{1}{2m}\hat{p}^2+
 \frac{1}{2}m\omega^2\hat{q}^2+ \frac{1}{3}\lambda\hat{q}^3\,.
\]
The cubic term could be seen as a perturbation for $\lambda$ sufficiently
small, provided $q$ does not grow too large.
Now, equations of motion read
\[
 \frac{{\rm d}}{{\rm d} t}\langle\hat{q}\rangle =
\frac{1}{m}
\langle\hat{p}\rangle \quad,\quad
 \frac{{\rm d}}{{\rm d} t}\langle\hat{p}\rangle =
-m\omega^2\langle\hat{q}\rangle-
\lambda\langle\hat{q}\rangle^2-(\Delta q)^2
=-V'(\langle\hat{q}\rangle)-\lambda(\Delta q)^2\,,
\]
coupling expectation values to the position fluctuation. (Similar equations
for $\hbar\to0$ have been used in \cite{Hepp} to prove that quantum mechanics
has the correct classical limit.)  The position fluctuation, in turn, obeys
\[
 \frac{{\rm d}}{{\rm d} t}(\Delta q)^2 =
\frac{2}{m}C_{qp}\quad,\quad
 \frac{{\rm d}}{{\rm d} t} C_{qp}= \frac{1}{m}(\Delta p)^2- m\omega^2(\Delta
q)^2+ 6\lambda \langle\hat{q}\rangle (\Delta q)^2
+3\lambda G^{0,3}
\]
and depends, via the evolution equation of $C_{qp}$, on a third-order moment
$G^{0,3}=\langle(\hat{q}-\langle\hat{q}\rangle)^3\rangle$ (called
skewness). Proceeding in this way, computing an equation of motion for
$G^{0,3}$ and so on, shows that infinitely many variables are coupled to one
another and to expectation values.

For a systematic formulation, we use, following \cite{EffAc,Karpacz}, the
quantum phase space of classical variables $q=\langle\hat{q}\rangle$ and
$p=\langle\hat{p}\rangle$ together with the moments
\begin{equation}
 G^{a,n}:=\langle(\hat{q}-\langle\hat{q}\rangle)^{n-a}
(\hat{p}-\langle\hat{p}\rangle)^a\rangle_{\rm Weyl}
\end{equation}
for $n\geq 2$, $a=0,\ldots,n$. (The subscript ``Weyl'' indicates that all
operator products are Weyl ordered before taking the expectation value,
averaging over all possible orderings.) For these variables we have Poisson
brackets
\begin{equation}
 \{\langle\hat{A}\rangle,\langle\hat{B}\rangle\}=
\frac{\langle[\hat{A},\hat{B}]\rangle}{i\hbar}
\end{equation}
extended by imposing the Leibniz rule to products of expectation values, as
they appear in moments. This general definition implies $ \{q,p\}=1$,
$\{q,G^{a,n}\}=0=\{p,G^{a,n}\}$ and a rather complicated (but explicitly
known) relation for $\{G^{a,n},G^{b,m}\}$ \cite{EffAc,HigherMoments}.

\subsection{Effective dynamics}

Evolution on the quantum phase space is determined by the quantum Hamiltonian
$H_Q(\langle\cdot\rangle,G^{\cdot,\cdot})=
\langle\hat{H}\rangle_{\langle\cdot\rangle,G^{\cdot,\cdot}}$, defined as a
function of expectation values and moments characterizing a state
used to compute the expectation value. The dynamical flow of the quantum
Hamiltonian couples expectation values and moments: the law
\begin{equation} \label{dOdt}
 \frac{{\rm d}\langle\hat{O}\rangle}{{\rm d}t} =
\frac{\langle[\hat{O},\hat{H}]\rangle}{i\hbar}=
\{\langle\hat{O}\rangle,\langle\hat{H}\rangle\}
\end{equation}
in general contains product terms multiplying expectation values and
moments. In this systematic way, quantum back-reaction is implemented.

A well-studied example illustrating all important features is the general
anharmonic oscillator, with classical Hamiltonian
$H=\frac{1}{2m}p^2+\frac{1}{2}m\omega^2q^2+U(q)$. If we first introduce
dimensionless variables $\tilde G^{a,n}=\hbar^{-n/2}(m\omega)^{n/2-a}G^{a,n}$,
we can compute the quantum Hamiltonian $H_Q:=\langle\hat{H}\rangle$ that
enters the Poisson bracket in (\ref{dOdt}), exhibiting all quantum corrections
with explicit factors of $\hbar$. By Taylor expansion, we have
\begin{eqnarray}
H_Q &=& \langle H(\hat{q},\hat{p})\rangle=\langle
H(\langle\hat{q}\rangle+(\hat{q}-\langle\hat{q}\rangle),
\langle\hat{p}\rangle+(\hat{p}-\langle\hat{p}\rangle))\rangle \\
&=&\frac{1}{2m}\langle\hat{p}\rangle^2+
\frac{1}{2}m\omega^2\langle\hat{q}\rangle^2+U(\langle\hat{q}\rangle)
+\frac{\hbar\omega}{2}(\tilde{G}^{0,2}+\tilde{G}^{2,2})
 +\sum_{n>2}\frac{1}{n!}
\left(\frac{\hbar}{m\omega}\right)^{n/2}U^{(n)}(\langle\hat{q}\rangle)\tilde{G}^{0,n}
\nonumber 
\end{eqnarray}
with the zero-point energy $\frac{1}{2}\hbar\omega
(\tilde{G}^{0,2}+\tilde{G}^{2,2})$ and a whole series of coupling terms.  The
series, in general, is asymptotic, as usual for semiclassical expansions. If
it is truncated to a finite sum up to $n_{\rm max}$, we obtain the
semiclassical approximation of order $n_{\rm max}$.

The quantum Hamiltonian $H_Q$ generates Hamiltonian equations of motion
$\dot{f}=\{f,H_Q\}$ on quantum phase space according to (\ref{dOdt}): from
\cite{EffAc}, 
\begin{eqnarray}
\dot{\langle\hat{q}\rangle}&=& \frac{\langle\hat{p}\rangle}{m}\label{eom}\\
\dot{\langle\hat{p}\rangle}&=&-m\omega^2\langle\hat{q}\rangle 
-U'(\langle\hat{q}\rangle)-\sum_n\frac{1}{n!}\left(
\frac{\hbar}{m\omega}\right)^{n/2}U^{(n+1)}(\langle\hat{q}\rangle)\tilde{G}^{0,n}\\
\dot{\tilde{G}}{}^{a,n}&=&-a\omega
\tilde{G}^{a-1,n}+(n-a)\omega \tilde{G}^{a+1,n}
-a\frac{U''(\langle\hat{q}\rangle)}{m\omega}\tilde{G}^{a-1,n}\\
\nonumber&&+
\frac{\sqrt{\hbar}aU'''(\langle\hat{q}\rangle)}{2(m\omega)^{3/2}}
\tilde{G}^{a-1,n-1}
\tilde{G}^{0,2}
+\frac{\hbar
  aU^{''''}(\langle\hat{q}\rangle)}{3!(m\omega)^2}
\tilde{G}^{a-1,n-1}\tilde{G}^{0,3} \label{Gdot} \\
\nonumber&&
-\frac{a}{2}\left(
\frac{\sqrt{\hbar}U'''(\langle\hat{q}\rangle)}{(m\omega)^{3/2}}
\tilde{G}^{a-1,n+1}+\frac{\hbar
U^{''''}(\langle\hat{q}\rangle)}{3(m\omega)^2}\tilde{G}^{a-1,n+2}\right)+\cdots
\end{eqnarray}
with infinitely many coupled equations for infinitely many variables, clearly
a system that in this generality is difficult to manage. Nevertheless, we can
see some general properties:
\begin{itemize}
\item Quantum corrections arise from back-reaction of fluctuations and higher
  moments (loop corrections in the language of quantum field theory) unless
  the system is harmonic with $U(q)=0$ (or ``free'').
\item State properties such as fluctuations are computed if we solve our
  equations, starting from initial conditions (the interacting vacuum). There
  is no need to assume properties of dynamical semiclassical states, which in
  other schemes are often based on ad-hoc choices such as Gaussians as the
  simplest peaked states.
\item The procedure is manageable if a free system is available as perturbative
  basis. The most general form of such a system is one with a linear dynamical
  algebra $[\hat{J}_i,\hat{J}_j]= \sum_k C_{ij}{}^k\hat{J}_k$ for a complete
  set of basic operators $\hat{J}_i$ that includes the Hamiltonian $\hat{H}$.
  With canonical basic variables, $\hat{H}$ must be quadratic for this
  condition to be realized. More general, non-canonical examples are known in
  quantum cosmology \cite{BouncePert}.
\end{itemize}

\begin{figure}
  \includegraphics[height=.4\textheight]{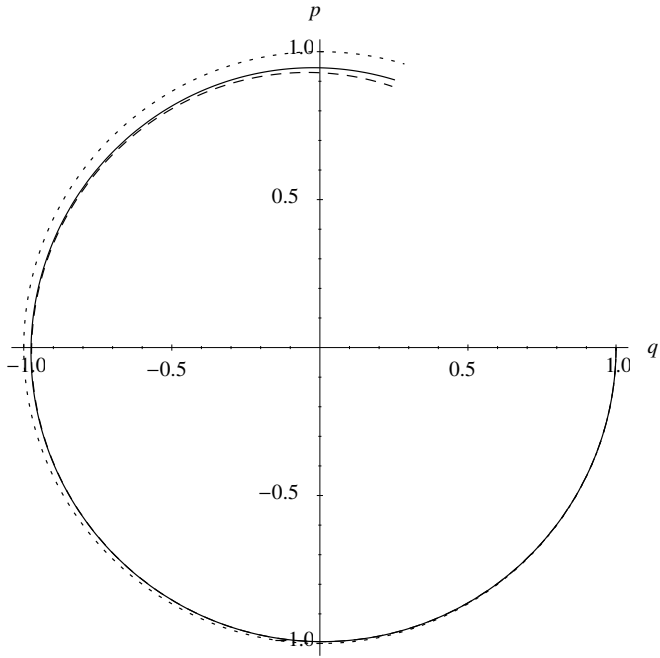}
  \caption{Position and momentum expectation values for the system shown in
    Fig.~\ref{RelOscWave}. Quantum back-reaction is captured to leading order
    in a semiclassical expansion as shown by the dashed curve
    \cite{EffConsComp}. Unlike the 
    classical orbit (dotted), the effective one follows the quantum
    corrections seen in the expectation value of the state starting
    semiclassically as in Fig.~\ref{RelOscWave}. (Initial
    values for the effective orbit are chosen so that they agree with
    expectation values and moments of the initial Gaussian profile of the
    semiclassical wave function.) \label{RelOscCircle}}
\end{figure}

\begin{figure}
  \includegraphics[height=.3\textheight]{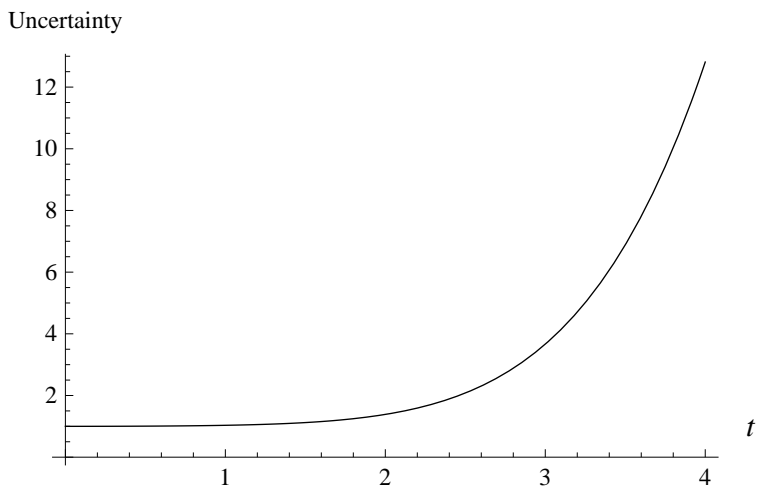}
  \caption{The uncertainty product $4\hbar^{-2}\left((\Delta q)^2(\Delta
      p)^2-C_{qp}^2\right)$, bounded from below by one according to
    (\ref{Uncert}), increases, showing that the state used in
    Figs.~\ref{RelOscWave} and \ref{RelOscCircle} evolves away from
    semiclassicality \cite{EffConsComp}. \label{RelOscUncert}}
\end{figure}

\begin{figure}
  \includegraphics[height=.2\textheight]{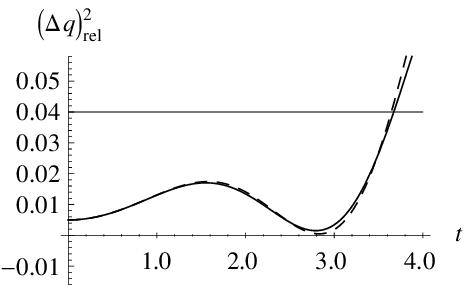}
  \includegraphics[height=.2\textheight]{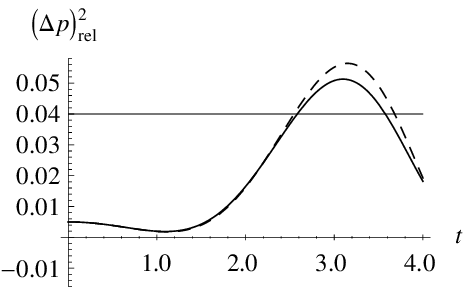}
  \includegraphics[height=.2\textheight]{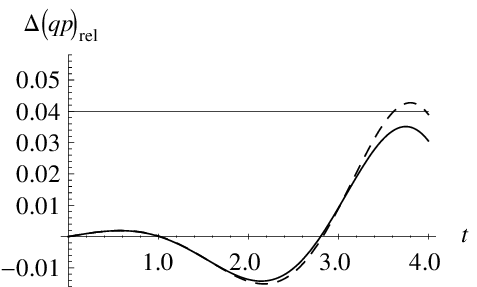}
  \caption{Relative second-order moments $\Delta(q^ap^b)_{\rm rel}= \Delta(
    q^ap^b)/\langle\hat{q}\rangle^a\langle\hat{p}\rangle^b$ 
   of the state shown in   Figs.~\ref{RelOscWave}  
    and \ref{RelOscCircle}, computed by effective equations (dashed) and from an
    evolving wave function (solid). When moments grow too large, surpassing
    the threshold indicated by the horizontal line, the semiclassical
    approximation to the order used breaks down \cite{EffConsComp}. 
    \label{RelOscMoments}}
\end{figure}

\begin{figure}
  \includegraphics[height=.3\textheight]{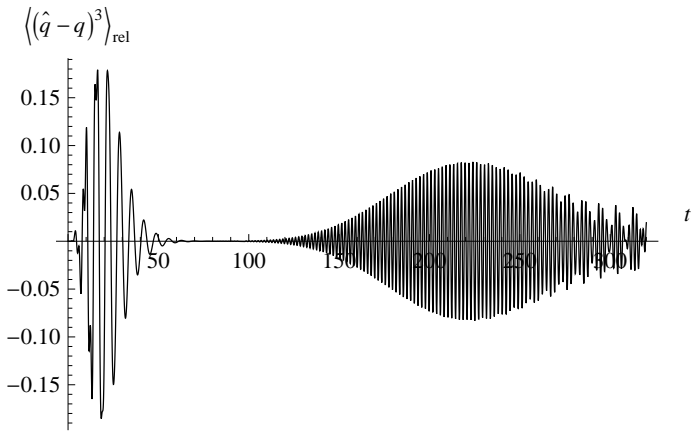}
  \caption{Long-term plot of the relative skewness $\Delta(q^3)_{\rm rel}=
    \Delta( q^3)/\langle\hat{q}\rangle^3$ for a wave
    function shown in Fig.~\ref{RelOscWave}. The state rapidly evolves away
    from a Gaussian, which would have zero skewness. Even though small values
    may again be attained later, the full state, characterized by infinitely
    many moments, may be far from being Gaussian
    \cite{EffConsComp}. \label{RelOscSkew}}
\end{figure}

\begin{figure}
\psfrag{ttdiv}{$t$}
\psfrag{Lambda}{}
\psfrag{LogGijPV}{even moments}
  \includegraphics[height=.25\textheight]{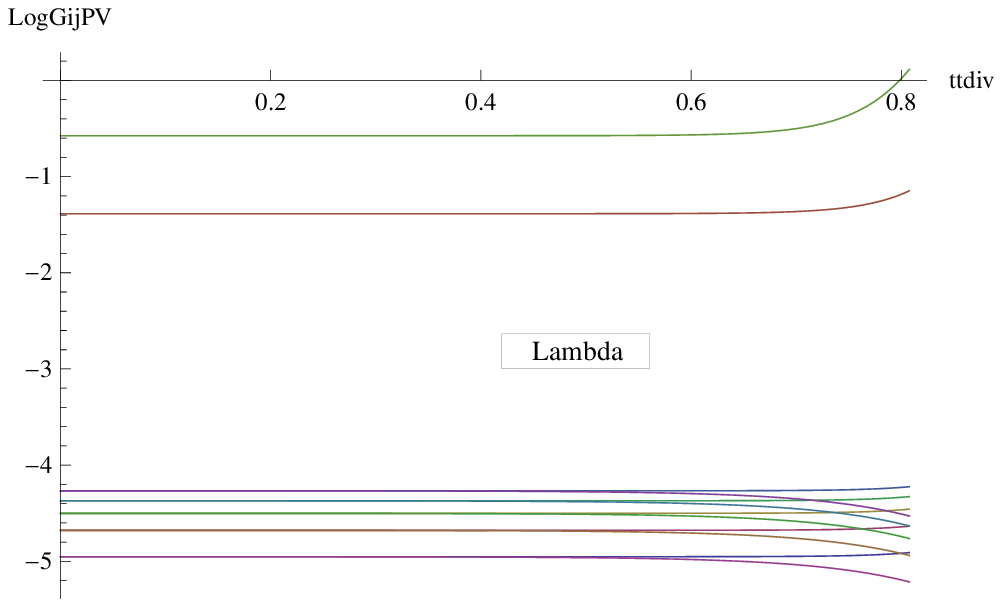}\hspace{0.5cm}
\psfrag{ttdiv}{$t$}
\psfrag{LogGij}{$\log(\mbox{odd moments})$}
\psfrag{Lambda}{}
   \includegraphics[height=.25\textheight]{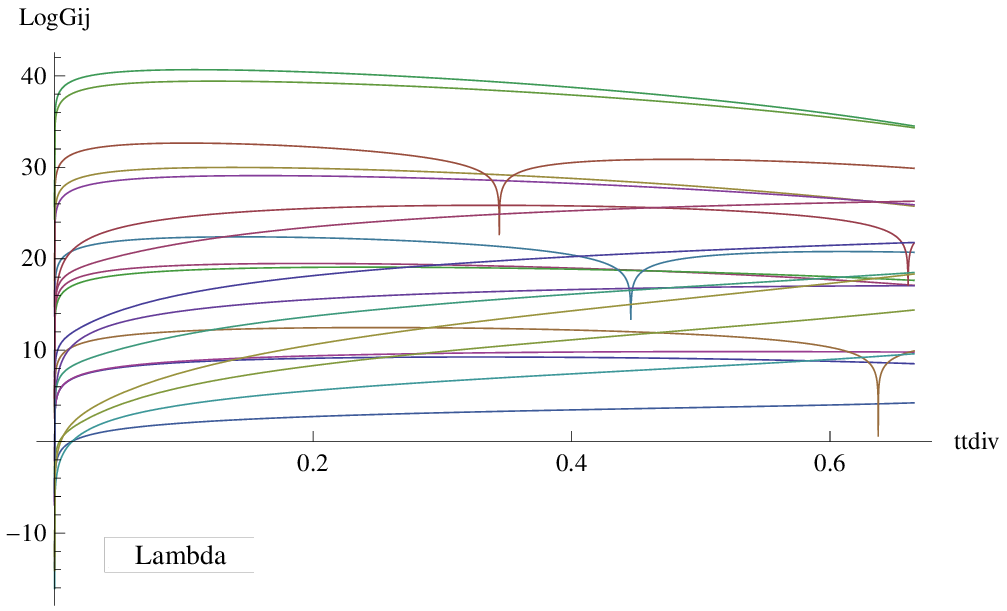}\hspace{0.5cm}
   \caption{Moments up to order 10, even at left and odd at right, for a
     cosmological model with a positive cosmological constant
     \cite{HigherMoments}. (Moments are rescaled by powers of $\hbar$
     depending on their order.) The state remains semiclassical for some time,
     as indicated by the hierarchy shown by the moments of different
     orders. However, although the initial state is Gaussian, with vanishing
     odd-order moments, the evolving state has a different, more complicated
     shape as shown by the non-zero odd-order moments on the right. Still, a
     dynamical hierarchy of the moments builds up as the non-Gaussian state is
     evolved.  \label{Higher}}
\end{figure}

Truncated to finite semiclassical order, the equations for expectation values
and moments are amenable to detailed numerical analysis; see
Figs.~\ref{RelOscCircle}--\ref{Higher} for examples. In many cases they can
also be solved analytically, at least if additional approximations are used to
decouple equations. An interesting expansion is obtained when the
semiclassical approximation is combined with an adiabatic one for the moments,
in which case one implements a derivative expansion.

We perform the adiabatic approximation by introducing a new (unphysical)
parameter $\lambda$, rescaling ${\rm d}/{\rm d}t$ to $\lambda {\rm d}/{\rm
  d}t$ in moment equations (\ref{Gdot}) and expanding $G^{a,n}=
\sum_eG_e^{a,n}\lambda^e$ \cite{EffAc}. We then solve equations order by order
in $\lambda$, thereby implementing the assumption of slow motion of the
moments. In the end, after the system has been solved, we set $\lambda=1$ as
required to recover the original equations. There is no guarantee
that $\lambda=1$ lies within the radius of convergence of the
expansion. Nevertheless, this form of the adiabatic expansion provides a
systematic way to arrive at a derivative expansion: higher orders in $\lambda$
introduce higher time derivatives \cite{HigherTime}.

To first order in $\hbar$ and zeroth in $\lambda$ we already obtain
interesting corrections. From (\ref{Gdot}), we then have the equation
\[
 0=\omega\left((n-a)G_0^{a+1,n}-a
 \left(1+\frac{U''(\langle\hat{q}\rangle)}{m\omega^2}\right)G_0^{a-1,n}\right) 
\]
for moments, which is algebraic instead of differential and has the general
solution
\begin{equation} \label{G0an}
 G_0^{a,n}={n/2 \choose a/2}{n \choose a}^{-1}\left(
 1+\frac{U''(\langle\hat{q}\rangle)}{m\omega^2}\right)^{a/2}G^{0,n}_0
\end{equation}
for even $a$ and $n$.  To this order, $G_0^{0,n}$ remains free.

To first order in $\lambda$, we must solve
\[
 (n-a)G_1^{a+1,n}-a\left(1+\frac{U''(\langle\hat{q}\rangle)}{m\omega^2}\right)
 G_1^{a-1,n}=\frac{1}{\omega}\dot{G}_0^{a,n}
\]
for $G_1^{a,n}$, but the equation also implies
\[
 \sum_{a\:{\rm even}}{n/2 \choose a/2}
\left(1+\frac{U''(\langle\hat{q}\rangle)}{m\omega^2}\right)^{(n-a)/2}
\dot{G}_0^{a,n}=0
\]
as a condition on moments (\ref{G0an}) of zeroth order in $\lambda$. They must
then be of the form
$G^{0,n}_0=C_n(1+U''(\langle\hat{q}\rangle)/m\omega^2)^{-n/4}$ with constants
$C_n$.  At this stage, all equations relevant to this order have been
implemented. The remaining freedom, the $C_n$, parameterize the choice of
specific states used. We have solved equations of motion (\ref{Gdot}) for
moments, corresponding to an evolving state starting with some initial wave
function. The initial state so far has not been restricted, and therefore
there must be free parameters left in our solutions of moments, exactly the
constants $C_n$. In the anharmonic case, we may fix the constants by stating
that the harmonic limit, $U(q)=0$, should bring us to the known moments of
some state of the harmonic oscillator, such as the ground state with
$C_n=2^{-n}n!/(n/2)!$. Using these values and inserting all solutions in our
equations of motion for expectation values, we obtain the zeroth
adiabatic-order correction \cite{EffAc}
\begin{eqnarray}
\dot{p}&=&-m\omega^2q -U'(q)-
\frac{\hbar}{2m\omega}U'''(q)G^{0,2}+\cdots\nonumber\\
&=&
-m\omega^2q -U'(q)-
\frac{\hbar}{4m\omega}\frac{U'''(q)}{\sqrt{1+U''(q)/m\omega^2}}+\cdots\,.
\end{eqnarray}

For comparison, we also mention the second adiabatic order whose derivation is
more lengthy \cite{EffAc}. As a second-order equation of motion, the canonical
effective description
\begin{eqnarray*}
&&\left(m+\frac{\hbar U'''(q)^2}{32m^2\omega^5\left(
 1+U''(q)/m\omega^2\right)
 ^{5/2}}\right)\ddot q\\
&&+\frac{\hbar\dot
 q^2\left(4m\omega^2U'''(q)U''''(q)\left(1+U''(q)/m\omega^2\right)-
 5U'''(q)^3\right)}
 {128m^3\omega^7\left(1+U''(q)/m\omega^2\right)^{7/2}}\\
&&+m\omega^2q+U'(q)+\frac{\hbar
U'''(q)}{4m\omega\left(1+U''(q)/m\omega^2\right)^{1/2}}=0\,.
\end{eqnarray*}
agress with results from the low-energy effective action \cite{EffAcQM}
\begin{eqnarray*}
\Gamma_{{\rm eff}}[q(t)]&=&\int
\md t\Biggl(\frac{1}{2}\left(m+\frac{\hbar U'''(q)^2}
{32m^2\omega^5\left(1+U''(q)/m\omega^2\right)^{5/2}}\right)\dot
q^2\\
&&
-\frac{1}{2}m\omega^2q^2-U(q)-\frac{\hbar\omega}{2} 
\left(1+\frac{U''(q)}{m\omega^2}
\right)^{1/2}\Biggr)\,.
\end{eqnarray*}
To higher orders in the adiabatic approximation, higher-derivative corrections
appear as well \cite{HigherTime}.

\subsection{Constrained systems}

With the methods described thus far, quantum corrections to canonical
Hamiltonian dynamics can be computed systematically, showing all instances of
state dependence. This approach is therefore useful for quantum gravity and
cosmology.  But gravity is a gauge theory and therefore requires the
implementation of constraints to remove spurious degrees of freedom. A the
same time, constraints generate gauge transformations corresponding, in this
case, to coordinate changes.

When quantized, following Dirac, classical constraints $C(q,p)=0$ on phase
space are turned into operator equations $C(\hat{q},\hat{p})|\psi\rangle=0$
for physical states. Alternatively, one may try to solve the classical
constraints and quantize the reduced phase space. Unfortunately, the resulting
phase spaces are often so complicated that known quantization techniques
cannot handle them. Moreover, important off-shell effects in quantum physics
may be overlooked: constraints arise as part of the system of equations of
motion, which should not be solved before quantum corrections have been
implemented. Quantization in general modifies the solution space.

Effective descriptions of constrained systems start with a construction
similar to effective Hamiltonian dynamics: We have a quantum constraint
$0=C_Q=\langle \hat{C}\rangle_{\langle\cdot\rangle, G^{\cdot,\cdot}}=C_{\rm
  class}(q,p)+\cdots$, expanded by moments, for every constraint operator
$\hat{C}$. However, this set of equations is not enough. A single constraint
in a first-class system on quantum phase space removes only two parameters
such as two expectation values, but not the corresponding moments. Just as a
classical pair of degrees of freedom becomes a whole tower of infinitely many
quantum degrees of freedom, we must have infinitely many constraints to
constrain them all. They can be obtained from the general expression
\cite{EffCons,EffConsRel}
\begin{equation}
 0=C_{f(q,p)}:=\langle f(\hat{q},\hat{p})\hat{C}\rangle_{\langle\cdot\rangle,
  G^{\cdot,\cdot}}
\end{equation}
with arbitrary phase-space functions $f(q,p)$.  Practically, polynomials
$f(q,p)$ suffice. To a given order in the moments, only finitely many $C_f$
then have to be considered.

Even at the effective level, quantum constraints and their solutions are
sensitive to issues normally dealt with as subtleties of Hilbert-space
constructions. After all, when we solve the quantum constraint equations, we
determine dynamical properties of expectation values and moments in physical
states. These variables are subject to requirements of unitarity, which often
is not automatic but must be implemented carefully. Effective constraints make
these procedures more manageable and systematic, compared to constructions of
physical Hilbert spaces for which only a few general construction ideas but
hardly any specific means exist. (Most examples use deparameterization without
providing a way to test for independence of the choice of time.)  We summarize
some of the salient features:
\begin{itemize}
\item The system of constraints, if there are more than one, is consistent and
  first class, provided the ordering $\langle
  f(\hat{q},\hat{p})\hat{C}\rangle$ is chosen. (All constraint equations are
  then left invariant under the flow generated by other constraints, when
  the constraints hold.) This ordering is not symmetric. 
\item As a consequence, the quantum constraint equations are not guaranteed to
  be real, and neither are their solutions. Indeed, we do not require reality
  of kinematical moments $G^{a,b}$ before the constraints are solved.
  Instead, we impose reality only after solving the quantum constraints to
  ensure physical normalization of states. The transition from complex-valued
  kinematical moments to real-valued physical ones, solving the constraints,
  corresponds to the transition from the kinematical Hilbert space ignoring
  the constraints to the physical Hilbert space free of gauge degrees of
  freedom. In this transition, the inner product, and therefore physical
  normalization, usually changes, especially when zero is contained in the
  continuous part of the spectrum of (some of) the constraints.
\item Different gauge fixings of the system of quantum constraints
  are related to different kinematical Hilbert-space structures. Again, the
  effective level provides more manageable techniques to describe different
  choices, with gauge transformations within one quantum phase space instead
  of unitary transformations between different Hilbert spaces. Such gauge
  transformations have, for instance, been made use of to help solve the
  problem of time in quantum gravity at least in semiclassical regimes
  \cite{EffTime,EffTimeLong,EffTimeCosmo}. These examples are the only ones in
  which it was possible to show that physical results in the quantum theory do
  not depend on one's choice of time.
\end{itemize}

Linear constraints provide the simplest examples, and even show, in spite of
their poor physical content, some interesting insights \cite{EffCons}. Let us
look at $\hat{C}=\hat{p}$, implying the quantum constraint
$C_Q=\langle\hat{p}\rangle=0$. The momentum expectation value is constrained
to vanish, while $\langle\hat{q}\rangle$ is pure gauge.  For second-order
moments, $C_p-C_Q^2=G^{pp}=(\Delta p)^2=0$ constrains the momentum
fluctuation, and $C_q=\langle\hat{q}\hat{p}\rangle=0$ the covariance. The
position fluctuation is pure gauge. To this order, therefore, all variables
--- expectation values and moments --- are eliminated, a pattern that extends
to all orders.

Another consequence is that the constraint
$C_q=\langle\hat{q}\hat{p}\rangle=0$ implies a complex-valued kinematical
covariance
\begin{equation}
 G^{qp}=\frac{1}{2}\langle\hat{q}\hat{p}+\hat{p}\hat{q}\rangle
-\langle\hat{q}\rangle\langle\hat{p}\rangle= -\frac{1}{2}i\hbar\,.
\end{equation}
With this solution, the uncertainty relation (\ref{Uncert}) is respected (and
saturated) even though one of the fluctuations vanishes. In this way, the
complex-valuedness of kinematical moments leads to overall consistency. In the
present example, no degree of freedom is left after all constraints have been
solved, and no reality conditions need be imposed. If there are additional,
unconstrained degrees of freedom, they can be restricted to be real, as
observables corresponding to expectation values and moments computed in the
physical Hilbert space.

\section{Application to canonical quantum gravity}

The techniques of the previous section provide the basis for an effective
theory of loop quantum gravity or, more generally, canonical quantum gravity.
In theories of gravity, we have several non-linear constraints with a
complicated algebra. Moreover, the constraints include the dynamics by a
Hamiltonian constraint and are therefore the most important part of those
theories. Unfortunately, no fully consistent quantization is known.  Effective
techniques again come in handy because they allow more manageable calculations
perturbative in $\hbar$ (or other expansions). This is sufficient for the
derivation of potentially observable phenomena.  Moreover, by analyzing
quantum corrections to the constraints and their algebra (which classically
exhibits the gauge structure of coordinate transformations) one can shed light
on modified space-time structures and even address fundamental questions.  

In the context of quantum gravity, the role of higher time derivatives,
alluded to before, becomes important. To recall, at adiabatic orders higher
than second, solutions for moments depend on higher time derivatives of
$\langle\hat{q}\rangle$. Higher-derivative effective actions then result.
Such terms are exactly what we expect in quantum gravity and cosmology, where
higher derivatives are part of higher-curvature terms. Non-effective
calculations directly in Hilbert spaces, on the other hand, have difficulties
making a connection with higher time derivatives.

Even in an effective setting, the correspondence is not entirely obvious, an
issue that once again is related to the question of general covariance.
Effective equations depend on the quantum state used, via initial values for
moment equations ${\rm d}G^{\cdot,\cdot}/{\rm d}t= \cdots$. There is no
gravitational Hamiltonian bounded from below, and therefore no obvious ground
state one might choose for an effective description as in the example of
anharmonic oscillators. And even if there were a ground state, it is not
clear at all if it would be a good choice. From the point of view of
non-perturbative quantum gravity, space-time in observationally accessible
regimes is in a highly excited state with huge expectation values of
geometrical operators such as the volume.

If a state $|0\rangle$ used to compute expectation values for an effective
description is Poincar\'e invariant (such as the Minkowski vacuum) and the
quantization in one's approach to quantum gravity is covariant, effective
constraints $\langle0|\hat{C}|0\rangle$ (or the effective action) are
covariant. However, we may not have a Poincar\'e invariant state in quantum
gravity; certainly a Minkowski vacuum as used in perturbative approaches would
not be fundamental. In such a situation, Poincar\'e transformations would not
be realized within one effective theory, even if the underlying
quantum-gravity theory is covariant. One would still be able to deal with
effective equations and obtain covariant results, somewhat analogous to
background-field methods. But more care would be required. Moreover, the usual
arguments for effective gravitational actions with nothing but
higher-curvature terms no longer hold: the setting is more general,
allowing for different quantum corrections, potentially stronger than
higher-curvature ones.

\subsection{Space-time}

\begin{figure}
  \includegraphics[height=.12\textheight]{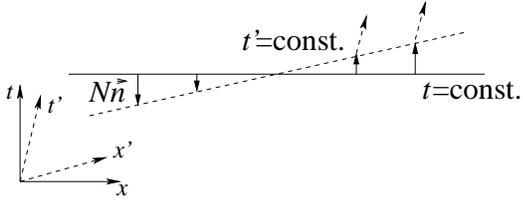}
  \caption{Space-time diagram with spatial slices related by a Lorentz
    boost. The result may be interpreted geometrically as a linear deformation
    of spatial slices by a function $N(x)$ along the normals. (Normals are
    drawn with right angles according to Minkowski geometry, looking
    non-perpendicular as drawn on a plane.)
 \label{Lorentz}}
\end{figure}

\begin{figure}
  \includegraphics[height=.2\textheight]{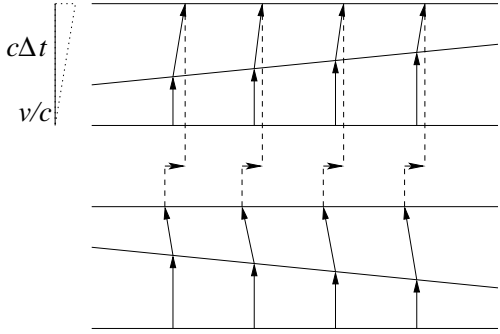}
  \caption{Normal deformations by $N_1(x)=v x/c$ (Lorentz boost) and
    $N_2(x)=c\Delta t-v x/c$ (reverse Lorentz boost and waiting $\Delta t$),
    performed in the two possible orderings (top and bottom). Geometry in the
    triangle shown implies that the two orderings lead to the same final
    slice, but with points displaced according to $\Delta x=v\Delta t$, in
    agreement with the commutator of a boost and a time translation. 
\label{HypDefLin}}
\end{figure}

To arrive at classifications of modified space-time structures, as they could
result from non-invariant quantum states, a geometrical representation of
space-time transformations is useful. We begin with a Lorentz boost of
velocity $v$,
\[
 x'=\frac{x-vt}{\sqrt{1-v^2/c^2}}\quad,\quad
 ct'=\frac{ct-vx/c}{\sqrt{1-v^2/c^2}}
\]
which implies a transformation of spatial slices $ct={\rm const}$ to $ct'={\rm
  const}$ in space-time. As shown in Fig.~\ref{Lorentz}, we may interpret
this transformation, as well as all other Poincar\'e transformations, as a 
linear deformation of the spatial slice, by distances
\[
 N(x)=c\Delta t+(v/c)\cdot x\quad,\quad
w(x)= \Delta x+{\bf R}\cdot x
\]
along the normal and within the slice, respectively. Also commutator relations
can be recovered geometrically by performing linear deformations in different
orderings; see Fig.~\ref{HypDefLin}. Similar geometrical relations are
obtained for all generators $P_{\mu}$ and $M_{\mu\nu}$ of the Poincar\'e
algebra
\begin{eqnarray}
  [P_{\mu},P_{\nu}] &=& 0\quad,\quad
  {} [M_{\mu\nu},P_{\rho}] = \eta_{\mu\rho}P_{\nu}-
  \eta_{\nu\rho}P_{\mu} \label{Poincare1}\\ 
  {} [M_{\mu\nu},M_{\rho\sigma}] &=& \eta_{\mu\rho}M_{\nu\sigma}-
  \eta_{\mu\sigma}M_{\nu\rho}- \eta_{\nu\rho}M_{\mu\sigma}+
  \eta_{\nu\sigma}M_{\mu\rho}  \label{Poincare2}
\end{eqnarray}

\begin{figure}
  \includegraphics[height=.13\textheight]{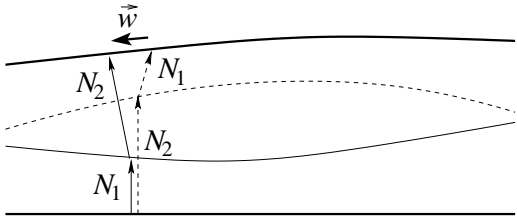}
  \caption{General relativity allows arbitrary coordinate transformations,
    geometrically implemented by non-linear deformations of spatial
    slices. Two normal deformations commute up to a spatial diffeomorphism by
    a vector field $\vec{w}$ according to (\ref{HH}). \label{SurfaceDef}}
\end{figure}

General relativity allows arbitrary coordinate changes, and thus non-linear
deformations of spatial slices. Again we obtain an algebra by performing
deformations in different orderings, as shown in Fig.~\ref{SurfaceDef} for two
normal deformations. We obtain the hypersurface-deformation algebra with
infinitely many generators $D[N^a]$ (tangential deformations along $N^a(x)$,
the spatial shift vector fields) and $H[N]$ (normal deformations by $N(x)$,
the lapse functions):
\begin{eqnarray}
 [D[N^a],D[M^a]]&=& D[{\cal L}_{M^a}N^a]\\
{} [H[N],D[M^a]] &=& H[{\cal L}_{M^a}N] \label{HypDef}\\
{} [H[N_1],H[N_2]] &=& D[q^{ab}(N_1\partial_bN_2-N_2\partial_bN_1)] \label{HH}
\end{eqnarray}
with the induced metric $q_{ab}$ on any spatial slice (and Lie derivatives in
commutators involving spatial deformations).

The hypersurface-deformation algebra is a natural extension of the Poincar\'e
algebra, which latter can be recovered by inserting linear functions $N=P_0+
x^a \tilde{M}_{a0}$ and $N_a= P_a+x^b\tilde{M}_{ba}$ for lapse and shift, with
coordinates $x^a$ referring to Minkowski space-time or at least a local
Minkowski patch. In addition to being infinite-dimensional, the
hypersurface-deformation algebra is much more unwieldy than the Poincar\'e
algebra. Both algebras depend on a metric, the Minkowski metric
$\eta_{\mu\nu}$ in (\ref{Poincare1}) and (\ref{Poincare2}), and the spatial
metric $q_{ab}$ in (\ref{HH}). However, while components of the Minkowski
metric are just constants, the spatial metric in the case of general
relativity depends on the position in space. Its appearance in (\ref{HH})
means that the algebra has not the usual structure constants, but structure
functions depending on an external coordinate which itself is not part of the
algebra. (Strictly speaking, the hypersurface-deformation algebra is not a Lie
algebra but a Lie algebroid, in rough terms a fiber bundle with a Lie-algebra
structure on its sections; see e.g.\ \cite{ConsAlgebroid}.) This feature of
the hypersurface-deformation algebra is responsible for many problems
associated with quantum gravity.

\subsection{Generally covariant gauge theory}

A generally covariant theory independent of the choice of coordinates on
space-time must be invariant under the hypersurface-deformation algebra, as a
more general, local version of the Poincar\'e algebra.  Since the induced
metric $q_{ab}$ changes under deformations of a spatial slice and appears in
structure functions, it is natural to take it as one of the canonical fields,
together with a momentum $p^{ab}$. On the resulting phase space, a gauge
theory is invariant under hypersurface deformations if there are constraints
$D[N^a]=0$ and $H[N]=0$ such that
\begin{eqnarray}
 \{D[N^a],D[M^a]\}&=& D[{\cal L}_{M^a}N^a]\\
 \{H[N],D[M^a]\} &=& H[{\cal L}_{M^a}N]\\
 \{H[N_1],H[N_2]\} &=& D[q^{ab}(N_1\partial_bN_2-N_2\partial_bN_1)]
\end{eqnarray}
is realized as an algebra under Poisson brackets. 

Any such theory is a generally covariant canonical theory of gravity
\cite{DiracHamGR}.  Space-time coordinate changes of phase-space functions
along vector fields $\xi^{\mu}= (\xi^0,\xi^a)$ are realized by the
Hamiltonian flow
\begin{equation}
 {\cal   L}_{\xi}f(q,p)= \{f(q,p), H[N\xi^0]+D[\xi^a+ N^a\xi^0]\}\,.
\end{equation}
(The additional coefficients of $N$ and $N^a$ result from a different
identification of directions in space-time and canonical formulations, the
former referring to coordinate directions, the latter to directions tangential
or normal to spatial slices; see \cite{LapseGauge,CUP}.)

Local invariance under hypersurface deformations is then equivalent to general
covariance, and an invariant theory in which hypersurface deformations are
consistently implemented as gauge transformations is the canonical analog of a
space-time scalar action.  Moreover, the symmetry is so strong that it
determines much of the dynamics: Hypersurface-deformation covariant
second-order equations of motion for $q_{ab}$ equal Einstein's equation
\cite{Regained,LagrangianRegained}.  All classical gravity actions, including
higher-curvature ones, have the same gauge-algebra (unless they break
covariance).

These important results leave only a few options for quantum
corrections. First, one may decide to break covariance. Since covariance is
implemented by gauge transformations, the theory is anomalous if the gauge is
broken.  Inconsistent dynamics results: the constraints $D[N^a]=0$ and
$H[N]=0$ are not preserved by evolution equations.  Inconsistency can formally
be avoided by fixing the gauge or frame before quantization, but this way out
does not produce reliable cosmological perturbation equations (see the
explicit example in \cite{GaugeInvTransPlanck}): Different choices of gauge
fixing within the same theory lead to different physical results after
quantization.  If the gauge is broken, the resulting quantum ``corrected''
theory is not consistent (unless there is a classically distinguished
frame). Breaking the gauge is widely recognized as a bad act to be avoided,
but still it often enters implicitly even in well-meaning approaches, most
often when deparameterization is used in quantum gravity.

The second option of quantum corrections is realized by approaches that
preserve the hypersurface-deformation algebra but allow equations of motion to
be of higher than second order, circumventing Hojman--Kucha\^r--Teitelboim
uniqueness of \cite{Regained,LagrangianRegained}.  We arrive at
higher-curvature effective actions.  Possible quantum corrections in cosmology
are then tiny, given by ratios of the quantum-gravity to the Hubble scale, or
$\rho/\rho_{\rm P}$ with the immense Planck density $\rho_{\rm P}$.

As the third option, we may allow for non-trivial consistent deformations of
the hypersurface-deformation algebra (and by implication the Poincar\'e
algebra). Full consistency is then realized because no gauge generator
disappears; only their algebraic relations change. Physically, we would obtain
quantum corrections in the space-time structure, not just in the dynamics, and
potentially new, not extremely suppressed corrections may result. This option
is not often considered, but it is realized in loop quantum gravity, where
\begin{equation} \label{HHbeta}
\{H_{(\beta)}[N_1],H_{(\beta)}[N_2]\}= D[\beta
q^{ab}(N_1\partial_bN_2-N_2\partial_bN_1)]
\end{equation}
with a phase-space function
$\beta$ implementing quantum corrections \cite{ConstraintAlgebra}.

Loop quantum gravity implies consistent deformations of the
hypersurface-deformation algebra. No gauge transformations are broken,
preserving consistency. As a consequence of the deformation, geometrical
notions may become non-standard. For instance, there is no effective line
element with a standard manifold because coordinate differentials in
\begin{equation}
 {\rm d}s^2_{\rm
    eff}=\tilde{g}_{ab} {\rm d}x^a{\rm d}x^b
\end{equation}
do not transform by deformed gauge transformations
$\{\cdot,H_{(\beta)}[N\xi^0]+D[\xi^a+N^a\xi^0]\}$ that change the
quantum-corrected spatial metric $\tilde{q}_{ab}$, usually completed
canonically to a space-time line element $-N^2{\rm d}t^2+ \tilde{q}_{ab} ({\rm
  d}x^a+N^a{\rm d}t)({\rm d}x^b+N^b{\rm d}t)$. Instead, one could try to use
non-commutative \cite{Connes} or fractional calculus \cite{Fractional} to
modify transformations of ${\rm d}x^a$, making ${\rm d}s_{\rm eff}^2$
invariant, but no such version has been found yet. Instead, once a consistent
algebra is known, one can evaluate the theory using observables according to
the deformed gauge algebra, for instance in cosmology
\cite{ScalarGaugeInv,LoopMuk,ScalarHol,ScalarTensorHol} or for black-hole
space-times \cite{ModifiedHorizon,ModCollapse,RNDeformed}. At this stage, {\em
  after} quantization, one may use gauge fixing of the deformed gauge
transformations or deparameterization because the consistency of the gauge
system with all its quantum corrections has been ensured.

\subsection{Loop quantum gravity}

To see how deformed constraint algebras and space-time structures arise in
loop quantum gravity, we should have a closer look at its technical
details. The basic canonical variables in this approach are the densitized
triad $E^a_i$ such that $E^a_iE^b_i= \det(q_{cd}) q^{ab}$, and the
Ashtekar--Barbero connection $A_a^i=\Gamma_a^i+\gamma K_a^i$ with the spin
connection $\Gamma_a^i$, extrinsic curvature $K_a^i$ and the Barbero--Immirzi
parameter $\gamma$ \cite{Immirzi,AshVarReell}. The canonical structure is
determined by
\[
 \{A_a^i(x),E^b_j(y)\}=
8\pi\gamma G\delta^b_a\delta^i_j\delta(x,y)\,.
\]

In preparation for quantization, one smears the basic fields by integrating
them to holonomies and fluxes,
\begin{equation}
 h_e(A)= {\cal P}\exp(\smallint_e  A_a^i\tau_i\dot{e}^a{\rm d}\lambda) 
\quad,\quad
 F_S(E)= \int_Sn_aE^a_i\tau^i{\rm d}^2y\,.
\end{equation}
The Poisson brackets of $A_a^i$ and $E^a_i$ imply a closed and linear
holonomy-flux algebra for $h_e(A)$ and $F_S(E)$, which is quantized by
representing it on a Hilbert space.  As a result, the Hilbert space is spanned
by graph states $\psi_{e_1,\ldots,e_n}(A)=f(h_{e_1}(A),\ldots,h_{e_n}(A))$
with curves $e_i$ in space, which are eigenstates of flux operators:
\begin{equation}
 \hat{F}_S\psi_{e_1,\ldots,e_n}\sim \ell_{\rm P}^2 {\rm
    Int}(S,e_1,\ldots,e_n)\psi_{e_1,\ldots,e_n}\,.
\end{equation}
Holonomies $h_e$ act as multiplication operators, creating spatial geometry in
two ways: (i) we may use operators for the same loop $e$ several times,
raising the excitation level per curve, or (ii) use different loops to
generate a mesh which, when fine enough, can resemble ordinary continuum
space.  Strong excitations are necessary for such a macroscopic geometry: loop
quantum gravity must deal with ``many-particle'' states.

Properties of the basic algebra of operators illustrate the discreteness of
spatial quantum geometry, and imply characteristic effects in composite ones,
such as the Hamiltonian constraint relevant for the dynamics.  Classically,
the constraints are polynomial in $A_a^i$. The quantized holonomy-flux algebra
provides operators $\hat{h}_e$, but none for ``$\hat{A}_a^i$.'' This feature
requires regularizations or modifications of the classical theory by adding
powers of $A_a^i$, completing the classical expression to a series of an
expanded exponential. Although there is a formal resemblance, these higher
orders are not identical to higher-curvature terms: they lack higher time
derivatives. As a second effect implied by flux operators $\hat{F}_S$ with
their discrete spectra, the theory has a state-dependent quantum-gravity scale
given by flux eigenvalues (which one may view as elementary areas). Depending
on the state, this scale may differ from the Planck scale if quantum geometry
is excited.  Dealing with a discrete version of quantum geometry, one must be
careful with potential violations of Poincar\'e symmetries
\cite{SmallLorentzViol}. The unbroken, deformed nature of quantum space-time
symmetries (\ref{HHbeta}) here provides consistency.

These general statements show that we should expect three types of corrections
in loop quantum gravity, irrespective of the detailed form of the theory.
First, as in all interacting theories, we have quantum back-reaction. In
gravitational theories, this is the key ingredient that provides
higher-derivative terms for curvature corrections
\cite{EffectiveGR,BurgessLivRev}. (For a related calculation in quantum
cosmology, see \cite{WdWCMB}.) The quantum structure of space then implies
additional corrections, not so much from the specific dynamics but from the
underlying quantum geometry. We have holonomy corrections, another form of
higher-order corrections with different powers of the connection.  In
cosmological regimes, these corrections are sensitive to the energy density,
just like higher-curvature corrections but in a different form. (Therefore,
these corrections should not play much of a role for potential observations.)
Finally, there are inverse-triad corrections that result from quantizing
inverse triads using the identity \cite{QSDI,QSDV}
\begin{equation} \label{Inverse}
 \left\{A_a^i,\int{\sqrt{|\det E|}}\mathrm{d}^3x\right\}= 2\pi\gamma G
 \epsilon^{ijk}\epsilon_{abc} \frac{E^b_jE^c_k}{{\sqrt{|\det E|}}}\,.
\end{equation}
The right-hand side is needed for the Hamiltonian constraint of gravity, but
flux operators are not invertible: they have discrete spectra containing
zero. The left-hand side, on the other hand, does not require an inverse
triad, and can be quantized using holonomy and volume operators, and turning
the Poisson bracket into a commutator divided by $i\hbar$. While the equation
is a classical identity, quantizing the left-hand side does not agree with
inserting triad (or flux) eigenvalues in the right-hand side: the third source
of quantum corrections \cite{InvScale}.

\subsubsection{Holonomy corrections: Signature change}

Holonomy corrections can easily be illustrated in isotropic models. With this
symmetry, connection variables are $A_a^i={c}\delta_a^i$ and
$E^b_j={p}\delta^b_j$ with $c=\gamma\dot{a}$ and $|p|=a^2$ depending only on
time. The Friedmann equation then reads
\begin{equation}
 -\frac{{c}^2}{\gamma^2|{p}|}+ \frac{8\pi G}{3}
 \rho=0\,.
\end{equation}
The use of holonomies in the quantum representation implies that there is no
operator for $c$ or $c^2$, but only one for any linear combination of
$\exp(i\mu c)$ with real $\mu$. To represent the Hamiltonian constraint
underlying the Friedmann equation, one therefore chooses a modification such
as
\begin{equation} \label{ModFried}
 \frac{{c}^2}{|{p}|}\mapsto \frac{\sin(\ell
 {c}/\sqrt{|{p}|})^2}{\ell^2}\sim \frac{{c}^2}{|{p}|}\left(1-
 \frac{1}{3}\ell^2\frac{{c}^2}{|{p}|}+\cdots\right)
\end{equation}
in terms of periodic functions, with some parameter $\ell$ related to the
precise quantization of the constraint.  If $\ell\sim \ell_{\rm P}$ is
Planckian, as often assumed, holonomy corrections are of the tiny order
$\ell_{\rm P}^2{c}^2/|{p}|\sim \rho/\rho_{\rm P}$ upon using the Friedmann
equation.

In isolation, holonomy corrections imply a ``bounce'' of isotropic
cosmological models, the main reason for interest in them. Writing the
constraint as a modified Friedmann equation,
\begin{equation} \label{ModFried}
 \frac{\sin(\ell {c}/\sqrt{|{p}|})^2}{\gamma^2\ell^2}= \frac{8\pi
   G}{3} \rho
\end{equation}
with a bounded left-hand side leads to an upper bound on the energy
density. Unlike classically, the density cannot grow beyond all bounds. At
this stage, however, the upper bound is introduced by hand, modifying the
classical dynamics (see also
\cite{RegularizationFRW,RegularizationLQC}). Although the modification is
motivated by quantum geometry via properties of holonomy operators, a robust
implementation of singularity resolution in this effective picture requires a
consistent implementation of quantum back-reaction and perturbative
inhomogeneity. Only if this complicated task can be completed can one claim
that a reliable quantum effect is realized, one that holds in the presence of
quantum interactions and takes into account correct quantum space-time
structures. In loop quantum cosmology \cite{LivRev,Springer}, this effective
picture has not yet been made robust, but there are more general
no-singularity statements based on properties of dynamical states
\cite{Sing,BSCG} and effective actions for them \cite{NoSing}. As we will see
below, the traditional bounce picture must be modified considerably when
inhomogeneity is taken into account.

Quantum back-reaction has been analyzed in bounce models by general effective
expansions \cite{BouncePot,QuantumBounce,BounceSqueezed} and numerically
\cite{HighDens}. While the equations remain complicated and not much is known
about solutions, it is clear that density bounds hold at least for
matter dominated by its kinetic energy term. The reason is that a free,
massless scalar, whose energy is purely kinetic, provides a harmonic model in
which basic operators and the Hamiltonian form a closed linear algebra (in a
suitable factor ordering) \cite{BouncePert}. No quantum back-reaction then
exists, and the model can be solved exactly. For kinetic-dominated matter, one
can use perturbation theory to show that bounds and bounces are still
realized, but without kinetic domination, for instance if there is a slow-roll
phase at high density, the presence of bounces remains questionable.

One should also note that even the presence of a bounce of expectation values
does not guarantee that evolution is fully deterministic. Especially in
harmonic cosmology, the evolution of fluctuations and some higher moments is
so sensitive to initial values that it is practically impossible to recover
the complete pre-bounce state from potentially observable information
afterwards \cite{BeforeBB,Harmonic}. (Claims to the contrary are based on
restricted classes of states.) This form of cosmic forgetfulness indicates
that the bounce regime does have unexpected features of strong quantum
effects, even when it is realized in a harmonic model free of quantum
back-reaction.

Quantum space-time structure in the presence of holonomy corrections implies
additional caveats, and finally removes deterministic trans-bounce
evolution. Quantum space-time structure follows from the
hypersurface-deformation algebra realized with holonomy corrections in the
presence of (at least) perturbative inhomogeneity. No complete version is
known, and it is not even clear if holonomy corrections can be fully
consistent. But some examples exist, in spherically symmetric models
\cite{JR,LTBII}, in $2+1$-dimensional models \cite{ThreeDeform} (with operator
rather than effective calculations) and for cosmological perturbations
ignoring higher-order terms in a derivative expansion of holonomies
\cite{ScalarHol}. In these cases, the hypersurface-deformation algebra is not
destroyed, implying consistency, but deformed: Instead of (\ref{HH}) we have
(\ref{HHbeta}).

In cosmological settings, the correction function for holonomies has the form
$\beta({c})=\cos(2\ell{c}/\sqrt{|{p}|})$ \cite{ScalarHol}.  Assuming maximum
density in (\ref{ModFried}) implies that $ \beta=-1$ is negative. (In the
linear limit of Fig.~\ref{HypDefLin}, we have the counter-intuitive
relation $\Delta x=-v\Delta t$ for motion.)  This sign change in the
hypersurface-deformation algebra implies that the space-time signature turns
Euclidean \cite{Action,SigChange} (to see this, one may draw
Fig.~\ref{HypDefLin} with Euclidean right angles for the normals), and indeed
evolution equations are elliptic rather than hyperbolic partial differential
equations \cite{ScalarHol}. (This is a concrete realization of the suggestions
in \cite{BohmEuclidean}, but by a different mechanism. It is also reminiscent
of the no-boundary proposal \cite{nobound}.)  There is no evolution through a
``bounce'', but rather a signature-change scenario of early-universe
cosmology. We obtain a non-singular beginning of Lorentzian expansion when
$\beta$ moves through zero depending on the energy density, a natural place to
pose initial values for instance for an inflaton state.

However, with uncertainties in quantum back-reaction and the precise form of
holonomy corrections, the deep quantum regime remains poorly controlled. There
is a significant amount of quantization ambiguities, and it remains unclear if
holonomy corrections can be fully consistent.  Higher-curvature and holonomy
corrections are both relevant at Planckian density, when $\rho\sim \rho_{\rm
  P}$, but they remain incompletely known.  Good perturbative behavior is
realized at observationally accessible densities far below the Planck density,
but the corrections are then so tiny that quantum gravity cannot be tested and
falsified based on them. Holonomy corrections, therefore, are not relevant for
potential observations.

\subsubsection{Inverse-triad corrections: Falsifiability}

Fortunately, loop quantum gravity offers a third option, inverse-triad
corrections, by which it becomes falsifiable. To illustrate their derivation,
we assume a lattice state with U(1)-holonomies $\hat{h}_e$ and fluxes
$\hat{F}_e$. (Normally, fluxes are associated with surfaces. But on a regular
lattice we can uniquely assign plaquettes to edges, so that an edge label for
fluxes is sufficient.) With one ambiguity parameter $0<r<1$, we then use
Poisson-bracket identities such as (\ref{Inverse}) to quantize an inverse flux
as
\begin{equation}
 \widehat{(|F|^{r-1} {\rm sgn}F)_e}=
 \frac{\hat{h}_e^{\dagger}|\hat{F}_e|^r\hat{h}_e-
   \hat{h}_e|\hat{F}_e|^r\hat{h}_e^{\dagger}}{8\pi Gr\gamma\ell_{\rm P}^2}
 =: \hat{I}_e\,.
\end{equation}
The relations $[\hat{h}_e,\hat{F}_e]= -4\pi\gamma\ell_{\rm
  P}^2\hat{h}_e$ and $\hat{h}_e\hat{h}_e^{\dagger}=1$, such that
\[
 \hat{h}_e^{\dagger}|\hat{F}_e|^r\hat{h}_e= |\hat{F}_e+4\pi\gamma\ell_{\rm
   P}^2|^r \quad,\quad \hat{h}_e|\hat{F}_e|^r\hat{h}_e^{\dagger}=
 |\hat{F}_e-4\pi\gamma\ell_{\rm P}^2|^r\,,
\]
allow us to compute the expectation value \cite{InflTest}
\begin{equation}
 \langle\hat{I}_e\rangle= {\frac{|\langle\hat{F}_e\rangle+4\pi\gamma\ell_{\rm
     P}^2|^r- |\langle\hat{F}_e\rangle-4\pi\gamma\ell_{\rm
     P}^2|^r}{8\pi Gr\gamma\ell_{\rm P}^2}}+ \mbox{moment terms}
\end{equation}
where we have already indicated a moment expansion as in effective equations.

\begin{figure}
  \includegraphics[height=.3\textheight]{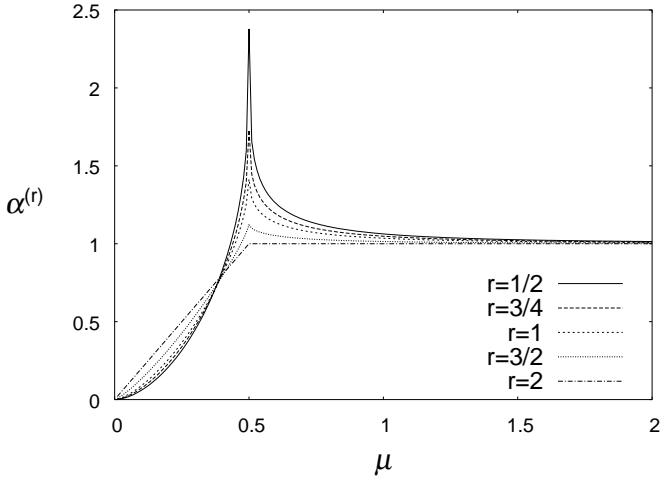}
  \caption{Inverse-triad correction function, depending on the ratio
    $\mu=L^2/(8\pi\gamma\ell_{\rm P}^2)$ of the discreteness scale $L$ and the
    Planck length. The parameter $r$ labels a form of quantization ambiguity,
    but does not change characteristic features \cite{QuantCorrPert}. 
\label{alpha}}
\end{figure}

We quantify these corrections, depending on a quantum-gravity scale
$\langle\hat{F}\rangle=:L^2$ related to flux expectation values, by a
correction function
\begin{equation} \label{alphaL}
\alpha(L):= \frac{\langle\hat{I}\rangle}{I_{\rm class}}=
 \frac{|L^2+4\pi\gamma\ell_{\rm P}^2|^r-
  |L^2-4\pi\gamma\ell_{\rm P}^2|^r}{8\pi\gamma r\ell_{\rm P}^2}L^{2(1-r)}
\end{equation}
whose $r$-dependent form is shown in Fig.~\ref{alpha}.

There are characteristic properties in spite of quantization ambiguities such
as the one parameterized by $r$ \cite{Ambig,ICGC}. For instance, for large
fluxes, the classical value $\alpha=1$ is always approached from above.
Inverse-triad corrections are large if the discreteness scale $L$ is nearly
Planckian, and small for larger vaues. The somewhat counter-intuitive feature
that inverse-triad corrections are large for small discreteness scale $L$ can
be understood from the fact that the corresponding operators eliminate
classical divergences at degenerate $E^a_i$, near $L=0$ \cite{InvScale}. In
this regime, for small fluxes, the corrections must therefore be strong. The
behavior then has a welcome consequence: Two-sided bounds on quantum
corrections.  Inverse-triad corrections are large for small $L$, and other
discretization effects for instance in the gradient terms of matter
Hamiltonians are large for large $L$. There is not just an upper bound, which
one could always evade by tuning parameters. The theory becomes falsifiable. A
theoretical estimate based on the relation to discretization and holonomy
effects shows that $\delta=\alpha-1>10^{-8}$ \cite{LoopMuk}.

A less welcome consequence of the dependence on $L$ (the state-dependent
lattice spacing) is that the size of corrections cannot easily be
estimated. But keeping $L$ as a parameter of the theory, its two-sided bounds
still allow the effects to be tested. Such a state dependence is to be
expected because effective equations depend on the state, for which there is
no natural choice, such as a vacuum, in quantum gravity. Also holonomy
corrections are subject to this ambiguity, even though it is often claimed
that they are completely determined by the energy density, a classical
parameter. However, this is the case only if the parameter $\ell$ in holonomy
modifications (\ref{ModFried}) is fixed, for instance by the popular but
ad-hoc choice $\ell\sim\ell_{\rm P}$, making holonomy corrections tiny and
irrelevant for potential observations. Since holonomy corrections and
inverse-triad corrections enter one and the same operator, the Hamiltonian
constraint, and must refer to the same state, there is only the parameter $L$
that determines their size. Holonomy corrections therefore are not uniquely
fixed, and if one were to declare that $L=\ell_{\rm P}$, inverse-triad
corrections would be large and the theory be ruled out. The freedom of
parameters cannot be avoided based on theoretical arguments alone.

As with holonomy corrections, the question of anomaly freedom and quantum
space-time structure must be addressed before corrections and their effects
can be taken seriously. Inverse-triad corrections are in a better position
than holonomy corrections, with a more-complete status of consistent
deformations. The Hamiltonian constraint is modified by the correction
function $\alpha$ from (\ref{alphaL}) multiplying each term that contains the
inverse triad classically:
\begin{eqnarray}
 H_{(\alpha)}[N] &=& \frac{1}{16\pi\gamma G} \int_{\Sigma} \mathrm{d}^3x 
N\alpha 
 \left(\epsilon_{ijk}F_{ab}^i\frac{E^a_jE^b_k}{\sqrt{|\det
E|}} +H_{\rm L} \right)\nonumber\\
&&+H^{(\alpha)}_{\rm matter}[N] +\mbox{``counterterms''}
\end{eqnarray}
where $F_{ab}^i$ is the curvature of the Ashtekar--Barbero connection, $H_L$
is an extra piece depending on extrinsic curvature, and ``counterterms'' are
determined by the condition of anomaly freedom \cite{ConstraintAlgebra}.
Together with the uncorrected diffeomorphism constraint
$D[N^a]=\int_{\Sigma}{\rm d}^3x D_aN^a= \int_{\Sigma}{\rm d}^3x
F_{ab}^iE^b_iN^a$, the algebra then reads
\begin{eqnarray}
 \{D[N^a],D[M^a]\}&=& D[{\cal L}_{M^a}N^a]\\
 \{H_{(\alpha)}[N],D[M^a]\} &=& H_{(\alpha)}[{\cal L}_{M^a}N]\\
 \{H_{(\alpha)}[N],H_{(\alpha)}[M]\} &=& 
D[\alpha^2 q^{ab}(N\partial_bM-M\partial_bN)]
\end{eqnarray}
(assuming constraints second order in inhomogeneity and $\delta= \alpha-1$
small). The same algebra has been found in spherical symmetry
\cite{LTBII,JR,ModCollapse} and in $2+1$-dimensional models using operator
calculations \cite{TwoPlusOneDef}. As with holonomy corrections, the form of
the deformation appears robust and universal. In the linear limit as in
Fig.~\ref{HypDefLin}, we have $\Delta x=\alpha^2 v\Delta t$ and therefore
expect quantum corrections to propagation. Unlike for holonomy corrections,
the algebra is always modified by a positive factor, and signature change does
not happen.

Corrections to propagation are realized more explicitly in Mukhanov-type
equations for gauge-invariant scalar and tensor perturbations \cite{LoopMuk},
\begin{equation}
 -u''+s^2\Delta u +(\tilde{z}''/\tilde{z})u=0\quad,\quad
 -w''+ \alpha^2\Delta w +(\tilde{a}''/\tilde{a})w=0
\end{equation}
with $s\not=\alpha$ (a known but lengthy function) and corrected
$\tilde{z}(a)$, $\tilde{a}(a)$.  The propagation speed differs from the speed
of light, and yet, as seen by the consistent constraint algebra, general
covariance is not broken but deformed.  A promising and rare feature is the
fact that different corrections are found for scalar and tensor modes:
corrections to the tensor-to-scalar ratio should be present, an interesting
aspect regarding potential observations. Like the corrections themselves,
effects are sensitive to the ratio $\langle\hat{F}\rangle/\ell_{\rm P}^2$, not
directly to the density. In contrast to $\rho/\rho_{\rm P}$, this parameter
can be significant during inflation. Observational evaluations
\cite{InflTest,InflConsist} indeed provide an upper bound on inverse-triad
corrections, which together with the theoretical lower bound allows only a
finite window $10^{-8}<\delta=\alpha-1<10^{-4}$ open by a reasonably small
number of orders of magnitude.

\section{Outlook}

It is not certain whether loop quantum gravity can be fundamental, and the
deep quantum regime (around the big bang) remains ambiguous and uncontrolled.
Nevertheless, the theory can be tested thanks to inverse-triad corrections,
with the following properties:
\begin{itemize}
\item Quantum effects are sensitive to the microscopic quantum-gravity scale
  in relation to the Planck length, not to the classical energy density as
  expected from higher-curvature corrections.  Therefore, they can be
  significant in observationally accessible regimes.
\item They provide a consistent deformation of the classical
  hypersurface-deformation algebra, and thereby a well-defined notion of
  quantum space-time.
\item There is a viable phenomenology thanks to a small number of
  low-curvature parameters.
\end{itemize}
All this is possible in the effective view of loop quantum gravity, in which
the main achievement of the theory is to provide new terms for effective
actions of quantum gravity. Long-standing conceptual issues, such as the
problem of time, can partially be solved at least in semiclassical regimes and
do not preclude progress in physical evaluations of the theory. A systematic
perturbation expansion is available in which  space-time structure and the
dynamics of observables are consistently treated in the same setting.

\section*{Acknowledgements}

I am grateful to the organizers of the Sixth International School on Field
Theory and Gravitation 2012 (Petr\'opolis, Brazil) for the invitation to give
a lecture series on which this write-up is based, and to the participants of
the school for interesting discussions and suggestions. Research discussed
here was supported in part by NSF grant PHY0748336.


\end{document}